\documentclass[preprint,prl
]{revtex4}

\usepackage{graphicx}
\usepackage{times}

\begin{document}
    
\title{Proposal for a simple quantum error correction test gate in 
linear optics}
\author{ T.C.Ralph}
\address{Centre for Quantum Computer Technology,
\\ Department of Physics,\\
University of Queensland, QLD 4072, Australia
\\Fax: +61 7 3365 1242  Telephone: +61 7 3365 3412 \\
email: ralph@physics.uq.edu.au\\}


\begin{abstract}
We describe a linear quantum optical circuit capable of demonstrating a 
simple quantum error correction code in a four photon experiment.
\end{abstract}
\maketitle 

\section{Introduction}

Whilst high hopes are held for the eventual demonstration of large 
scale quantum processing, present experimental attention remains 
fixed on few qubit demonstrations. Implementation of few qubit 
algorithms in the competing platforms provides valuable insight into 
the important physics and technical issues of different architectures.

A relatively new contender for scalable quantum computation is the 
scheme due to Knill et al \cite{KLM} based on single photon, dual 
rail qubits; linear optical networks; and photon resolving measurement and 
feedforward. Simple gates based on this scheme have been demonstrated 
\cite{pit02} and more general gates are planned \cite{ral01,pit01,ral02}. 
It is thus timely to consider what small scale circuits might be 
possible with the currently available technology.

A key enabling quantum circuit is error correction \cite{shor95,ste96}. 
Even medium scale quantum processing is expected to be impossible without error 
correction. Here we describe a simple example of error 
correction which would suit demonstration with a linear optical circuit. 
In its simplest form the experiment would require only three coincident photons.

\section{The Circuit}

The quantum circuit we wish to consider is a simplification of the 
standard bit flip correcting code \cite{nie00} and 
is shown in Fig.1.  The qubit 
is encoded with the following relationship between logical, $|- 
\rangle_{L}$, and physical, $|- \rangle$, qubits: 
\begin{eqnarray}
    |0 \rangle_{L} & = & |0 \rangle |0 \rangle \nonumber\\
    |1 \rangle_{L} & = & |1 \rangle |1 \rangle 
\end{eqnarray}
We suppose that one of the physical qubits, say the second, suffers 
decoherence which produces random bit-flips. As a result of this 
decoherence an arbitrary initial 
qubit: 
\begin{eqnarray}
\alpha |0 \rangle_{L} + \beta |1 \rangle_{L} = \alpha |0 \rangle |0 
\rangle + \beta |1 \rangle |1 \rangle
\label{encst}
\end{eqnarray}
evolves into the mixed state:
\begin{eqnarray}
\rho & = & (1-P)(\alpha |0 \rangle |0 
\rangle + \beta |1 \rangle |1 \rangle) (\alpha^{*} \langle 0 | \langle 0 
| + \beta^{*} \langle 1 | \langle 1 |) + \nonumber\\
 &  & P(\alpha |0 \rangle |1 
\rangle + \beta |1 \rangle |0 \rangle) (\alpha^{*} \langle 0 | \langle 1 
| + \beta^{*} \langle 1 | \langle 0 |)
 \label{error state}
\end{eqnarray}
where $P$ is the probability that the bit flip occurred. The aim of 
the circuit which follows is to return the qubit to its original 
logical value. To achieve this requires an ancilla photon prepared in 
the physical zero state. We can write the combined state of 
the qubits and ancilla after decoherence as
\begin{figure}[ht]
     \begin{center}
     \includegraphics[width=15cm]{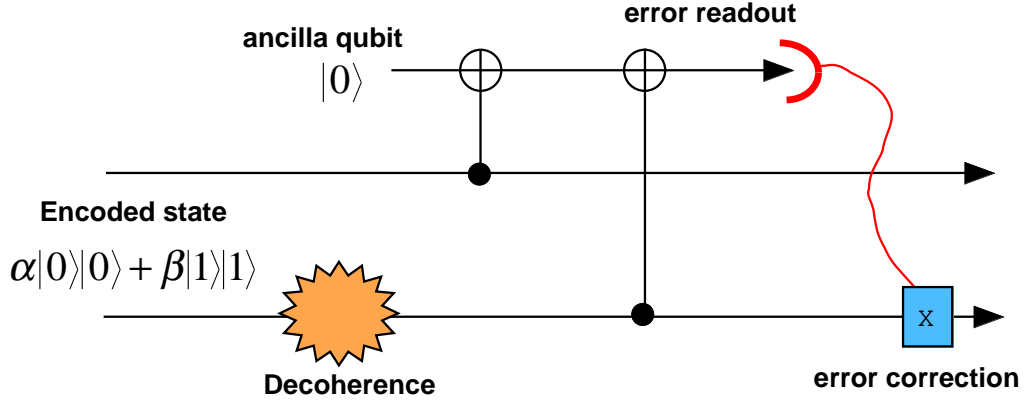}
     \end{center}
     \vspace{-5mm} \caption{Schematic of simple error corection code.}
     \label{Expt}
\end{figure}
\begin{eqnarray}
\rho & = & (1-P)|0 \rangle_{a} (\alpha |0 \rangle |0 
\rangle + \beta |1 \rangle |1 \rangle ) 
(\alpha^{*} \langle 0 | \langle 0 
| + \beta^{*} \langle 1 | \langle 1 |) \langle 0 |_{a} 
+ \nonumber\\
 &  & P |0 \rangle_{a} (\alpha |0 \rangle |1 
\rangle + \beta |1 \rangle |0 \rangle) 
(\alpha^{*} \langle 0 | \langle 1 
| + \beta^{*} \langle 1 | \langle 0 |) \langle 0 |_{a}
 \label{error state+}
\end{eqnarray}
with the subscript ``a'' labelling the ancilla state. 
A controlled not (CNOT) gate is applied with the ancilla as target and the 
first qubit as control. This transforms the state to 
\begin{eqnarray}
\rho & = & (1-P)(\alpha |0 \rangle |0 
\rangle |0 \rangle_{a}+ \beta |1 \rangle |1 \rangle |1 \rangle_{a}) 
(\alpha^{*} \langle 0 | \langle 0 
| \langle 0 |_{a} + \beta^{*} \langle 1 | \langle 1 | \langle 1 |_{a}) 
+ \nonumber\\
 &  & P(\alpha |0 \rangle |1 
\rangle |0 \rangle_{a} + \beta |1 \rangle |0 \rangle |1 \rangle_{a}) 
(\alpha^{*} \langle 0 | \langle 1 
| \langle 0 |_{a} + \beta^{*} \langle 1 | \langle 0 | \langle 1 |_{a})
 \label{error state++}
\end{eqnarray}
A second CNOT is then applied with the ancilla as target but now the 
second qubit acts as control. The state becomes
\begin{eqnarray}
\rho & = & (1-P)|0 \rangle_{a} (\alpha |0 \rangle |0 
\rangle + \beta |1 \rangle |1 \rangle ) 
(\alpha^{*} \langle 0 | \langle 0 
| + \beta^{*} \langle 1 | \langle 1 |) \langle 0 |_{a} 
+ \nonumber\\
 &  & P |1 \rangle_{a} (\alpha |0 \rangle |1 
\rangle + \beta |1 \rangle |0 \rangle) 
(\alpha^{*} \langle 0 | \langle 1 
| + \beta^{*} \langle 1 | \langle 0 |) \langle 1 |_{a}
 \label{error state+++}
\end{eqnarray}
Finally we detect the ancilla state. If we find the ancilla in the 
zero state then the logical qubit is projected onto its original state and no 
correction is neccessary. On the other hand if the ancilla is found 
in the one state then the projected state is 
\begin{eqnarray}
\alpha |0 \rangle |1 
\rangle + \beta |1 \rangle |0 \rangle
\end{eqnarray}
We know an error has occurred which can be corrected by flipping the 
value of the second physical qubit and thus returning the logical 
qubit to its initial value.

The circuit can also be understood in the language of stabilizer 
codes \cite{got96}. The code space is the $+1$ eigenstates of $ZZ$, 
whilst the error space is the $-1$ eigenstates of $ZZ$. Here $Z$ is 
the Pauli sigma z operator and $ZZ$ indicates the tensor product of a 
sigma z measurement on the first qubit with a sigma z measurement on 
the second qubit. The two CNOT's achieve precisely this measurement 
with the ancilla equals zero result indicating the $+1$ eigenstate, and 
hence the system being in the code space, while the ancilla equals 
one result indicates the $-1$ eigenstate and hence the system is in 
the error space and needs to be corrected.

The usefulness of this circuit is obviously limited by the very 
specific nature of the errors corrected, ie only bit flips on one of 
the physical qubits. Never-the-less it exhibits the same basic structure 
as more versatile codes \cite{nie00} 
whilst limiting the complexity of the required 
circuit. In the next section we will discuss an optical implementation 
of this code which appears tractable to current experimentation.

\section{The Optical Implementation}

We consider single photon qubits with the polarization degree of 
freedom determining their logical values. We define the physical qubit 
value ``zero'' as being a single horizontally polarized photon, $| 0 \rangle 
\equiv | H \rangle$, and a physical qubit value of ``one'' as being a 
single vertically polarized photon, $| 1 \rangle 
\equiv | V \rangle$. Knill et al \cite{KLM} showed that 
non-deterministic CNOT gates could be constructed from linear optics  
with success rates of one in sixteen, using two additional ancilla 
photons. A CNOT would be required to produce the encoded logical 
qubits. Thus an implementation of our error correction circuit would 
require three CNOTs and hence require the simultaneous production of 
nine single photon states. The success rate would be about one in four 
thousand. Although not beyond the realm of medium term possibility, 
such an experiment is currently not feasible. However in the following 
we will discuss an in principle demonstration utilizing the 
coincidence basis with much lower technical requirements.
\begin{figure}[ht]
     \begin{center}
     \includegraphics[width=15cm]{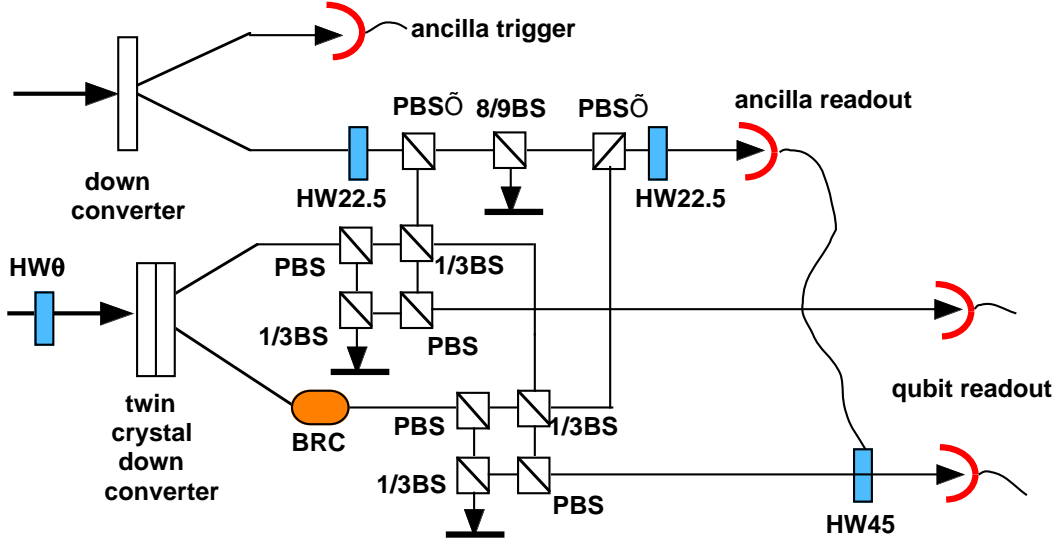}
     \end{center}
     \vspace{-5mm} \caption{Schematic of experimental proposal. 
     PBS: HV polarizing beamsplitter; PBS': VH polarizing beamsplitter; 
     $\eta$BS: beamsplitter of reflectivity $\eta$; BRC: birefringent 
     crystal; HW$\theta$: half-wave plate with rotation angle $\theta$.}
     \label{Expt}
\end{figure}
The proposed set up is shown schematically in Fig.2. The encoded state 
is produced directly using type one down conversion through a pair 
of $\chi 2$ crystals with 
orthogonally oriented optical axes \cite{whi99}. Pairs of photons 
originating from one crystal will be horizontally polarized whilst 
those originating from the other crystal will be vertically polarized. 
By changing the polarization of the pump beam the proportion of 
horizontal to vertical pairs can be continuously varied. In the far 
field, for sufficiently thin crystals, spatial information on the 
origin of the pairs is erased and the resulting entangled output state is 
approximately
\begin{eqnarray}
|vac \rangle_{1}|vac \rangle_{2}+\chi(\alpha |H \rangle_{1} |H 
\rangle_{2} + \beta |V \rangle_{1} |V \rangle_{2})
\end{eqnarray}
where $|vac \rangle_{i}$ is the vacuum state, $1$ and $2$, label the 
two beams, $\alpha=\cos\theta$ and $\beta=\sqrt{1-\alpha^{2}}$ and 
$\theta$ is the orientation of the pump beam polarization away from 
vertical. Coincidence detection will pick out only the doubly 
occupied parts of the state, so the effective input state is
\begin{eqnarray}
\alpha |H \rangle_{1} |H 
\rangle_{2} + \beta |V \rangle_{1} |V \rangle_{2}
\end{eqnarray}
which is logically identical to Eq.\ref{encst}. 

Controlled decoherence can be introduced onto the second qubit by 
passing it through a birefringent crystal oriented at $45$ degrees to 
horizontal \cite{jam01}. The effect of the crystal is to pull apart in 
time the two polarization modes in the diagonal/anti-diagonal basis. 
When the time difference becomes an appreciable fraction of the 
coherence length decoherence occurs. In the horizontal/vertical basis 
the resut is random bit flips and the state produced can be written in 
the form of Eq.\ref{error state}. The degree of decoherence, $P$, is a 
function of the crystal length.

To detect the errors we must introduce an additional ancilla qubit 
for readout. This can be supplied by a second down-converter producing 
just horizontal pairs. A CNOT gate which works in the coincidence 
basis and does not require additional ancilla modes 
can be implemented using just linear optics \cite{ral02,hof02}. The 
schematic of Fig.2 shows the optical network needed to implement the 
two CNOT's in the quantum circuit (Fig.1). The polarization qubits 
are decomposed into separate spatial modes using polarizing 
beamsplitters then the modes from different qubits are mixed on 
beamsplitters. Many 
optical paths are possible through the network but only a few 
result in photonic qubits at all three outputs, as determined by coincident 
detection of photons. In such cases quantum interference due to 
indistinguishability of photons ensures the required transformations are 
implemented. The 
success rate is one in eight-one.

Correction of the decohered qubit could be implemented, if required by 
the result of the ancilla detection, using a fast Pockel cell
\cite{pit02b}. 

For most runs of the experiment insufficient photons will be detected 
and we have a null result. However, on those occasions when a photon 
is detected at the ancilla trigger and at the ancilla output, {\it 
and} photons are detected at both qubit outputs, then to an excellent 
approximation the quantum circuit of Fig.1 will have been 
implemented. For such event we would expect the logical value of 
the qubits should ideally be the same as that prepared, in spite of 
the presence of the decohering element. 

Four photon coincidences of a 
few per minute have been achieved with down conversion \cite{bou02}. 
This will be 
further reduced by a factor of one in eighty-one for this proposal. 
On the other hand 
working in the coincidence basis means, at least in principle, that 
this source efficiency and the efficiency of the detectors does not 
effect the fidelity of the accepted events.

\section{Conclusion}

We have described a simple error correction code, and proposed an in 
principle test of its operation using current four-photon technology. 
Although both the code and its implementation are major 
simplifications over what would be required in a scalable 
architecture, many of the basic principles are common. We thus suggest 
that pursuit of experiments like the one proposed here will reveal 
much about the important physical and technical issues to be faced for 
truly scalable architectures.

\acknowledgements

We thank Charlene Ahn and Gerard Milburn for motivating discussions. 
This work was supported by the Australian Research Council and ARDA.

\end{document}